\documentstyle[aps,preprint,pra,epsfig]{revtex}

\begin{document}

\draft

\title{Propagation of Discrete Solitons in Inhomogeneous Networks}

\author{R. Burioni$^1$, D. Cassi$^1$, P. Sodano$^2$, 
A. Trombettoni$^1$, and A. Vezzani$^1$}
\address{$^1$ I.N.F.M. and Dipartimento di Fisica, Universit\`a
di Parma, parco Area delle Scienze 7A Parma, I-43100, Italy}
\address{$^2$ Dipartimento di Fisica and Sezione I.N.F.N.,
Universit\`a di Perugia, Via A. Pascoli Perugia, I-06123, Italy}

\date{\today}
\maketitle

\begin{abstract} 
In many physical applications solitons propagate on supports
whose topological properties may induce new and interesting effects. In this
paper, we investigate the propagation of solitons on chains with a topological 
inhomogeneity generated by the insertion of a 
finite discrete network on the chain. 
For networks connected by a link to a single site of the
chain, we derive a general criterion 
yielding the momenta for perfect reflection and transmission 
of traveling solitons and we discuss solitonic motion on chains 
with topological inhomogeneities. 
\end{abstract}

%\pacs{PACS: 74.25.Dw, 05.30.Jp, 74.50.+r, 85.25.Cp}

%\begin{multicols}{2}
 
{\bf In the last decades, a huge amount of work has been devoted 
to the study of the propagation of discrete solitons in regular, 
translational invariant lattices. 
However, in several systems, like networks of 
nonlinear waveguide arrays, 
Bose-Einstein condensates in optical lattices, 
arrays of superconducting Josephson junctions and silicon-based 
photonic crystals, one can engineer the shape 
(i.e., the topology) of the network. Correspondingly, 
an interesting task is
the study of the propagation of solitons in inhomogeneous networks. 
The general idea of this work 
is that network topology strongly affects
the soliton propagation. We provide a 
general argument giving the momenta of perfect transmission and 
reflection for a soliton scattering through a finite general network  
attached to a site of a chain: the momenta of perfect transmission and 
reflection are related in a simple way to the energy levels of the 
attached network. This criterion directly links the transmission coefficients 
with the network adjacency matrix, which encodes all the
relevant informations on its topology. Such relation puts into evidence 
the topological effects on the soliton propagation. 
The situations where finite linear chains, Cayley trees and other 
simple structures are attached to a site of an unbranched chain are 
investigated in detail.}

\section{Introduction}  

The analysis of nonlinear models on regular lattices 
\cite{flach98,scott99,hennig99}, 
as well as the investigation of linear models on 
inhomogeneous and fractal networks \cite{nakayama94} 
has attracted a great deal of attention in the last decades: 
while nonlinearity dramatically modifies the dynamics, allowing for 
soliton propagation, energy localization, and the existence of discrete
breathers \cite{scott99}, topology mainly affects the
energy spectrum giving rise to interesting phenomena such as anomalous
diffusion, localized states, and fracton dynamics \cite{nakayama94}. 
It is now both timely and highly desirable to begin a thorough 
investigation of nonlinear models on general inhomogeneous networks, 
since one expects not only interesting 
new phenomena arising from the interplay 
between nonlinearity and topology, but also an high potential impact 
for applications to biology \cite{peyrard89,special} and 
to signal propagation in optical waveguides \cite{kivshar03}. 
Recently, the effects of uniformity break on soliton propagation 
\cite{christodoulides01,kevrekidis03} 
and localized modes \cite{mcgurn02} has been investigated 
by considering $Y$-junctions  
\cite{christodoulides01,kevrekidis03} (consisting of a  
long chain inserted on a site of a chain yielding 
a star-like geometry) or geometries like  
junctions of two infinite waveguides or the waveguide coupler 
\cite{mcgurn02}. Here, we consider general finite 
networks inserted on a chain.

The discrete nonlinear Schr\"odinger equation (DNLSE) is a paradigmatic 
example of a nonlinear equation on a lattice which has been 
successfully applied to several contexts \cite{kevrekidis01,ablowitz04}: 
in particular, it 
has been used to describe the physics of 
arrays of coupled optical waveguides \cite{eisenberg98,morandotti99} 
and arrays of Bose-Einstein condensates \cite{trombettoni01}. 
It is well known that, on a homogeneous chain, 
the DNLSE is not integrable \cite{ablowitz04}; nevertheless,
soliton-like wavepackets can propagate for a long time and 
the stability conditions of soliton-like solutions can be derived 
within variational approaches \cite{malomed96}. 
Furthermore, the dynamics of traveling pulses has been investigated 
in detail in literature \cite{duncan93,flach99,gomez04} (more references 
are in \cite{kevrekidis01}). 
The simplest example of an inhomogeneous 
chain is provided by an external potential localized on a site of the chain 
\cite{forinash94,konotop96,krolikowski96,aceves96,kevrekidis01}:  
an experimental set up with a single defect has been recently 
realized with coupled optical waveguides \cite{mandelik03}. 
Another relevant example of an inhomogeneous chain is obtained 
adding an additional Fano degree of freedom coupled to 
the a site of the chain, which gives the so-called Fano-Anderson 
model (see \cite{miroshnichenko03,flach03,miroshnichenko05} 
and references therein).

As a first step in the investigation of the properties of nonlinear models 
on general inhomogeneous networks, we shall analyze the 
propagation of DNLSE solitons on a class 
of inhomogeneous networks built by suitably adding a 
finite topological inhomogeneity to an unbranched chain. 
The general framework where this analysis can 
be carried out is provided by graph theory \cite{harary69}; 
in particular, we shall consider networks where a finite 
discrete graph $G^0$ is attached by a link to a site of the 
homogeneous, unbranched chain (see Fig.1) 
while all the sites potentials $\epsilon_i$ are set to 
a constant. Such systems may be experimentally realized 
by placing the nonlinear waveguides in a suitable 
inhomogeneous arrangement, like the one depicted in Fig.1. We mention 
that in arrays of Bose-Einstein condensates one can build up geometries, 
which differ from the unbranched chain, by properly superimposing 
the laser beams creating the optical lattices \cite{oberthaler03}. 
In superconducting Josephson arrays \cite{fazio01}, 
present-day technologies allow for to prepare the insulating support for 
the junctions in order to create structures 
of the form "chain + a topological defect".
In the context of coupled nonlinear waveguides \cite{hennig99}, 
one should couple the waveguides according the geometry 
of the graph $G^0$, and couple this network of waveguides to a single 
waveguide of the unbranched chain; a similar engineering should be 
requested to realize photonic crystal circuits \cite{birner01} 
obtained merging the circuit $G^0$ to the unbranched chain.

As we shall discuss in Section IV, 
the shape of the attached graph $G^0$ affects the transmission
and reflection coefficients as a function of the soliton momentum: 
as an example of this general phenomenon, we consider 
unbranched chains to which simple graphs, like finite chains and Cayley trees, 
are added. Our analysis points to the fact that 
the topology of the network (i.e., how its sites are connected) controls 
the transmission properties, and that one can modify soliton propagation 
by varying the topology of the inserted network. In particular, we shall 
show that the momenta of perfect transmission are determined by the energy 
levels of the inserted graph, i.e., by the
eigenfrequencies of the $G^0$'s oscillation modes in the linear case 
(see Section IV).   
    
On a chain, stable solitonic wavepackets can propagate for long times 
\cite{kevrekidis01}. When a graph $G^0$ is inserted, one can study 
how the presence of this topological inhomogeneity modifies the soliton 
propagation. We numerically evaluate the transmission 
coefficients and we compare the numerical results 
with analytical findings obtained for a relevant soliton class, 
to which we refer as {\em large-fast solitons}. 
For large fast solitons the transmission coefficients
can be evaluated within a linear approximation \cite{miroshnichenko03}. 
Indeed the characteristic time for the soliton-topological defect collision
are very small with respect to the soliton dispersion 
time; therefore, the soliton scattering can be approximated by
the scattering of a plane-wave. 
However, as we numerically checked in the Figs.2-6, 
the nonlinearity still plays a role, giving long-lived solitons, especially 
near the momenta of perfect reflection or perfect transmission: 
it keeps the soliton shape during its propagation. 
Since in many experimental settings 
one can easily check if the reflected wavepacket 
is vanishing, this work could provide a basis for a topological 
engineering of solitonic propagation on inhomogeneous networks. 

The plan of the paper is the following. In Section II we review the
properties of the DNLSE on a chain and we introduce the variational
approach to investigate the soliton dynamics. 
In Section III, by using graph theory \cite{harary69}, we
define the DNLSE on inhomogeneous networks built by adding a 
topological perturbation to an unbranched chain; furthermore, we explain the
numerical techniques used in the paper for the study of the soliton
scattering, and we discuss the range of validity of the linear approximation
used in the analytical computations. In Section IV we present 
our analysis yielding
the conditions on the spectrum of the finite graph $G^0$ in 
order to obtain total reflection and transmission of solitons. 
In Section V we study the relevant case when $G^0$
is a finite, linear chain and we show that the Fano-Anderson model 
\cite{miroshnichenko03,flach03} can be realized within our approach 
by considering a single link attached to the unbranched chain. 
In Section VI we show that, for self-similar graphs $G^0$, the values 
of momenta for which perfect reflection occurs becomes perfect 
transmission momenta when the next generation of the graph is considered and 
we study in detail the case of Cayley trees as an 
example of self-similar structures. In Section VII we study the transmission 
coefficients for three different inserted finite graphs: 
loops, stars and complete graphs. Finally,  
Section VIII is devoted to our concluding remarks. 

\section{DNLSE on a chain}

Besides its theoretical interest, 
the DNLSE describes the properties of interesting systems, such as 
arrays of coupled optical waveguides and arrays
of Bose-Einstein condensates. On a chain the DNLSE reads
\begin{equation}
\label{DNLS-retta}
i  \frac{\partial \psi_n}{\partial \tau} = - \frac{1}{2} (\psi_{n+1} 
+ \psi_{n-1}) + \Lambda \mid \psi_n \mid ^2 \psi_n + \epsilon_{n} \psi_n
\end{equation}
where $n$ is an integer index denoting the site position and
the normalization condition is $\sum_n \mid \psi_n \mid ^2=1$. 
In Eq.(\ref{DNLS-retta}) one has a kinetic coupling term only between 
nearest-neighbour sites, but the effect of next- nearest-neighbour 
coupling and long-term coupling has been also often considered 
(see the reviews \cite{hennig99,kevrekidis01} for more references): 
in the present paper we consider only a constant 
nearest-neighbour interaction, but from the next Section 
we allow for that the number of nearest-neighbours of a site 
is not constant across the network (like for the simple chain), but it can 
vary according the topology of the graph.
 
In condensate arrays, $\psi_n(\tau)$ is the wavefunction of the 
condensate  in the $n$th well. Time $\tau$ is in 
units of $\hbar/2K$, 
where $K$ is the tunneling rate between neighbouring condensates; 
$\epsilon_n=E_n/2K$ where $E_n$ 
is an external on-site field superimposed to the optical lattice and 
the nonlinear coefficient is $\Lambda=U/2K$, where $U$ 
is due to the interatomic interaction 
and it is proportional to the scattering length ($U$ 
is positive for $^{87}Rb$ atoms and is negative for $^{7}Li$ atoms). 

In arrays of one-dimensional coupled optical waveguides
\cite{eisenberg98} $\psi_n(\tau)$ is the electric 
field in the $n$th-waveguide at the position $\tau$ and the DNLSE
describes the spatial evolution of the field.
The parameter $\Lambda$ is proportional to the Kerr nonlinearity 
and the on-site potentials $\epsilon_n$ are
the effective refraction indices of the individual waveguides.
As the light propagates along the array, the coupling
induces an exchange of power among the single waveguides.
In the low power limit (i.e. when the nonlinearity is negligible),
the optical field spreads over the whole array.
Upon increasing the power, the output field narrows
until it is localized in a few waveguides, and discrete solitons can 
finally be observed \cite{eisenberg98,morandotti99}.
Experiments with defects (i.e., with particular 
waveguides different from the others) have been already reported 
\cite{mandelik03}.  

On a chain DNLSE soliton-like wavepackets 
can propagate for a long time even if the equation 
is not integrable \cite{hennig99}. 
Let us consider, at $\tau=0$, a gaussian wavepacket centered 
in $\xi(\tau=0) \equiv \xi_0$, 
with initial momentum $k$ and width $\gamma(\tau=0) \equiv \gamma_0$: 
its time  
dynamics are studied resorting to the Dirac time-dependent 
variational approach \cite{dirac30} which well 
reproduces the exact results in the continuum theory \cite{cooper93}. 
In its discrete version the wavefunction can be written 
as a generalized gaussian
\begin{equation}
\label{gaussol}
\psi_{n}(\tau)=\sqrt{{\cal K}} \, \cdot \, 
e^{ -\frac{(n-\xi)^2}{\gamma^2} + ik(n-\xi) +
i \frac{\delta}{2}(n-\xi)^2}
\end{equation}
where $\xi(\tau)$ and $\gamma(\tau)$ are, respectively,
the center and the width of the density $\rho_n = \mid \psi_n \mid^2$,
and $k(\tau)$ and $\delta(\tau)$  are the momenta conjugate to 
$\xi(\tau)$ and $\gamma(\tau)$ respectively; 
${\cal K}$ is just a normalization factor.
The wave packet dynamical evolution is obtained from the Lagrangian 
${\cal L}= \sum_n  i \dot{\psi}_n \psi_n^\ast - \cal{H}$, 
with the equations of motion 
for the variational parameters $\xi,\gamma,k,\delta$. 
In the absence of external potential ($\epsilon_n=0$), 
one obtains the Lagrangian \cite{trombettoni01a}
\begin{displaymath}
{\cal L}={\cal K} \sum\limits_{n=-\infty}^{\infty} 
e^{-(2n^2+2n-4n\xi+2\xi^2-2\xi+1)/\gamma^2} 
\cos{[\delta(n+1/2-\xi)+k]} - 
\frac{\Lambda {\cal K}^2}{2} \sum\limits_{n=-\infty}^{\infty} 
e^{ - 4 (n-\xi)^2 / \gamma^2}  
\end{displaymath}
\begin{equation}
\label{LAG-N-finito}
+{\cal K} \sum\limits_{n=-\infty}^{\infty} \left\{
-\frac{\dot{\delta}}{2} (n-\xi)^2 + \delta \dot{\xi}(n-\xi) -
\dot{k}(n-\xi)+k \dot{\xi} \right\} \,  
e^{-2(n-\xi)^2/\gamma^2}. 
\end{equation}
With $\gamma$ not too small ($\gamma \gg 1$), we can replace 
the sums over $n$ with integrals: to evaluate the error committed, 
we recall that \cite{knuth89} 
\begin{equation}
\frac{\sum\limits_{n=-\infty}^{\infty} e^{-\frac{(n-\xi)^2}
{\gamma^2}}}{\int\limits_{-\infty}^{\infty} dn \, \, \, 
e^{-\frac{(n-\xi)^2}{\gamma^2}} } 
= 1+ O(e^{- \pi^2 \gamma^2}).
\label{knuth}
\end{equation}
In this limit the normalization factor becomes 
${\cal K}=\sqrt{2/\pi\gamma^2}$. 
We finally get \cite{trombettoni01} 
\begin{equation} 
\label{LAG}
{\cal L}=k \dot{\xi} - \frac{\gamma^2 \dot{\delta}}{8} - 
\frac{\Lambda}{2 \sqrt{\pi \gamma^2} } + \cos{k} \, \cdot \, 
e^{-\eta}, 
\end{equation}
where $\eta = 1 / 2 \gamma^2 + \gamma^2 \delta^2 / 8$.
The equations of motion are
\begin{eqnarray}
\dot{k} & = & 0 \label{var1} \\
\dot{\xi} & = & \sin{k} \, \cdot \, e^{-\eta} \label{var2} \\ 
\dot{\delta} & = & \cos{k} \Big(4 / \gamma^4-\delta^2 \Big) 
e^{-\eta} + 2 \Lambda / \sqrt{\pi} \gamma^3 \label{var3} \\ 
\dot{\gamma} & = & \gamma \delta \cos{k} \, \cdot \, 
e^{-\eta}: \label{var4}
\end{eqnarray} 
$k(\tau)=k$ is conserved. 
Notice that, due to the discreteness, the group velocity 
cannot be arbitrarily large ($\dot{\xi} \approx \sin{k} \le 1$). 
As Eq.(\ref{knuth}) clearly shows, the variational equations 
of motions (\ref{var4}) are meaningful only for large solitons: 
the Peierls-Nabarro potential does not appear in (\ref{var4}), 
and the equations feature momentum conservation. We mention 
that in uniform DNLSE chains a threshold condition for 
the soliton propagation appears: only if the soliton is
sufficiently broad solitons may freely move \cite{papa03}. In the following 
we shall consider only large-fast solitons, so that the variational 
equations of motions (\ref{var4}) are appropriate; however, to study 
the propagation of localized discrete breathers in inhomogeneous networks 
one should study the Lagrangian (\ref{LAG-N-finito}).

When $\dot{\gamma}=0$ and $\dot{\delta}=0$, the shape of the wavefunction
does not vary and one has a variational soliton-like solution 
where the center of mass move with a constant velocity $\dot{\xi}= constant$. 
If $\Lambda>0$ 
the conditions $\dot{\gamma}=0$ and $\dot{\delta}=0$ can be satisfied
only if $\cos{k}<0$ (i.e., only when the effective 
mass is negative): for this reason in the following 
we take only momenta $\pi/2 \le k \le \pi$ (positive velocities) 
or $-\pi \le k \le - \pi/2$ (negative velocities). In particular, 
for $\delta(\tau=0) \equiv \delta_0=0$ and large enough solitons 
($\gamma_0 \gg 1$), the condition on
$\Lambda$ allowing for a soliton solution is \cite{trombettoni01}
\begin{equation}
\label{lambda_sol}
\Lambda_{sol} \approx  2 \sqrt{\pi} \frac{\mid \cos{k} \mid} 
{\gamma_0}.
\end{equation}

The stability of variational solutions has been 
numerically checked showing that the shape of the solitons
is preserved for long times.
In the following, we use the term ``solitons'' to name the solutions of the 
variational equations (\ref{var1})-(\ref{var4}). 
One can have a similar criterion using other 
variational approaches \cite{malomed96}. One also expects that 
the integrable version of the DNLSE, the so-called Ablowitz-Ladik 
equation \cite{ablowitz76,ablowitz04}, 
provides results very similar to those obtained in this paper. 
%In the following, for a soliton solution, we will omit 
%the subscript $_0$ in $k$ and $\gamma_0$. 

\section{DNLSE on graphs}

The DNLSE (\ref{DNLS-retta}) can be generalized to a general discrete network
by means of graph theory. A graph $G$ is given by a set of sites $i$ 
connected pairwise by set of unoriented links $(i,j)$
defining a neighbouring relation between the sites. The topology 
of a graph is 
described by its adjacency matrix $A_{i,j}$
which is defined to be $1$ if $i$ and $j$ are nearest-neighbours, and $0$
otherwise. The DNLSE on a graph reads as 
\begin{equation} 
\label{DNLS-gen} 
i \frac{\partial \psi_i}{\partial \tau} = - \frac{1}{2} \sum_{j}  A_{i,j}
\psi_{j}+ \Lambda \mid \psi_i \mid ^2 \psi_i + \epsilon_{i} \psi_i
\end{equation}   

Equation (\ref{DNLS-gen}) describes the wavefunction dynamics in a wide range 
of discrete physical systems, and it can be applied to regular
lattices as well as to inhomogeneous networks such as fractals, 
complex biological structures and glasses. The properties of 
Eq.(\ref{DNLS-gen}) on small graphs has been investigated in 
\cite{eilbeck85}. 
The first term in the right-hand side of Eq.(\ref{DNLS-gen}) represents 
the hopping between nearest-neighbours (with tunneling rate proportional 
to $A_{i,j}$), the second is the nonlinear term, and the third one 
describes superimposed external potentials. 
We remark that in Eq.(\ref{DNLS-gen}) the numbers of nearest neighbours 
is site-dependent. Furthermore, since $A_{i,j}=1$ if $i$ and $j$ 
are nearest-neighbours, one is assuming that the tunneling rate 
between neighbouring sites entering Eq.(\ref{DNLS-gen}) 
is constant across the array and
it is (in the chosen units) equal to $1$. Below, we shall consider also 
the case of a tunneling rate which is not constant across the array. 

We shall focus on the situation where the graph $G$ is obtained 
by attaching a finite graph $G^0$ to a single site of the 
unbranched chain (see Fig.1) and setting
$\epsilon_i=0$ for all the sites. We denote the sites of the unbranched 
chain and of the
graph $G^0$ with latin indices $m,n,\dots$ and greek indices 
$\alpha, \beta, \cdots$ respectively. 
A single link connects the site $n=0$ of the chain with
the site  $\alpha$  of the graph $G^0$.

The scattering of a soliton through the topological perturbation 
can be numerically studied as follows. 
At $\tau=0$ (hereafter, we refer to $\tau$ as a time
even if for the optical waveguides it represents a spatial variable) one 
prepares a gaussian soliton (\ref{gaussol}) centered well to the left of $0$ 
(i.e., $\xi_0<0$) moving
towards $n=0$ ($\sin(k)>0$) with a width related to the  nonlinear 
coefficient according to Eq.(\ref{lambda_sol}). From Eq.(\ref{DNLS-gen}) 
the time evolution of the wavefunction may be numerically evaluated: 
when $\tau_s \approx \xi_0 /\sin(k)$ the soliton scatters through the 
finite graph $G^0$
($\sin(k)$ being the group velocity of the soliton). 
At a time $\tau$ well after
the soliton scattering (i.e. $\tau \gg \tau_s$), 
the reflection and transmission coefficients ${\cal R}$
and ${\cal T}$ are given by
\begin{equation} 
{\cal R}=\sum_{n<0} \mid \psi_n(\tau)\mid^2 \label{R}
\end{equation} 
\begin{equation} 
{\cal T}=\sum_{n>0} \mid \psi_n(\tau)\mid^2.
\label{T} 
\end{equation} 

Note that, while in the linear case ($\Lambda=0$) one has  ${\cal
R}+{\cal T}=1$, in general, nonlinearity violates unitarity by allowing for 
phenomena such as soliton trapping; nevertheless, there are regimes 
where soliton trapping is negligible and ${\cal R}+{\cal T} \approx 1$.
We numerically checked that in the time dynamics reported 
in the paper this condition is well satisfied.
Situations corresponding to resonant scattering (i.e. 
${\cal R}=0$ or ${\cal T}=0$) have a particular relevance: in fact, 
these situations can be easily experimentally detected, and the
soliton-like solution is stable also well after the scattering, 
as it is numerically verified 
in different examples of resonant reflection and transmission.

For an important class of soliton solutions (to which we 
refer as large-fast solitons) the scattering through a 
topological inhomogeneity can be analytically studied using a 
linear approximation.
The interaction between the soliton and the topological 
inhomogeneity is characterized by two time-scales: 
the time of the soliton-defect interaction  
$\tau_{int}=\gamma/\sin{k}$ and the soliton dispersion 
time (i.e. the time scale in which 
the wavepacket will spread in absence of interaction) $\tau_{disp}= 
\gamma/ (4 \sin{(1/2\gamma) \cos{k})}$ \cite{miroshnichenko03}. 
For {\em large} ($\gamma \gg 1$, as in many 
relevant experimental settings) and {\em fast} solitons, i.e.
\begin{equation}
\label{linearaprox}
v=\sin{k} \gg (2/ \gamma) \cos{k},
\end{equation}
one has that the soliton 
may be considered as a set of non interacting plane waves while 
experiencing scattering on the graph; thus the 
soliton transmission may be studied by considering, in the linear regime 
(i.e., $\Lambda=0$), the transport coefficients of a plane wave across the 
topological defect. The use of the linear 
approximation for the analysis of the interaction of a 
fast soliton with a local defect in the continuous 
nonlinear Schr\"odinger equation is reported in \cite{cao95}. 
Later, we shall compare the analytical findings with a numerical solution 
of Eq.(\ref{DNLS-gen}), namely with the reflection and 
transmission coefficients ${\cal R}$ and ${\cal T}$ given by 
Eqs.(\ref{R})-(\ref{T}).

\section{A general argument for resonant transmission}

In this Section we show that, if a large-fast soliton scatters through 
a topological perturbation of an unbranched chain, the soliton momenta 
for perfect reflection and transmission are completely determined 
by the spectral properties of the attached graph $G^0$: 
in particular one has ${\cal R}=1$ if 
$2 \cdot \cos{k}$ coincides with an energy level of $G^0$, while
${\cal T}=1$ if  $2 \cdot \cos{k}$ is an energy level 
of the reduced graph $G^r$, i.e., of the graph obtained from $G^0$ by 
cutting the site $\alpha$ from $G^0$ (see Fig.1).
In algebraic graph theory (see e.g. \cite{harary69,biggs74})
the energy level of a graph is simply defined as an eigenvalue of its
adjacency matrix. We will call $A^0$ and $A^r$ the adjacency matrices of
$G^0$ and $G^r$, respectively.

%More formally, the adjacency matrix ${ A }^0_{\eta,\eta'}$ 
%of the attached graph $G^0$ is 
%${ A }^0_{\eta,\eta'}=1$ when the sites $\eta$ and $\eta'$ 
%are nearest-neighbours of $G^0$ 
%and $0$ otherwise, and similarly for 
%${ A }_{\eta,\eta'}^r$ 
%(we recall that instead $A_{i,j}$ is the adjacency matrix of the whole 
%network). 

For large-fast solitons, 
the pertinent eigenvalue equation to investigate is   
\begin{equation}
\label{DLS}
-\frac{1}{2} \sum_{j} A_{i,j} \psi_j=\mu \psi_i
\end{equation}
(here $i$ is a generic site of the network). The solution 
corresponding to a plane wave 
coming from the left of the chain is $\psi_n=a e^{ikn}+be^{-ikn}$ 
for $n<0$ and $\psi_n=c e^{ikn}$ for $n>0$, so that 
$\mu=-\cos{k}$. The reflection coefficient is given by 
${\cal R}=\mid b / a \mid ^2$ and the transmission coefficient 
by ${\cal T}=\mid c / a \mid ^2$. 
The continuity at $0$ requires $a+b=c$. The equation in $0$ is
\begin{equation}
-\frac{1}{2} (a e^{-ik}+b e^{ik}+c e^{ik}+\psi_{\alpha})=-\cos{k} 
\cdot (a+b)
\label{eq_0}
\end{equation}
while in $\alpha$ one has
\begin{equation}
-\frac{1}{2} (a+b+\sum_{\eta \in G^0} A^0_{\alpha,\eta} \psi_\eta)=
-\cos{k} \cdot \psi_{\alpha}. 
\label{eq_alpha}
\end{equation}
At the sites $\eta$ of $G^r$ one obtains 
\begin{equation}
-\frac{1}{2} \sum_{\eta' \in G^0} A^0_{\eta,\eta'} \psi_{\eta'}=-\cos{k} 
\cdot \psi_\eta.
\label{eq_altri}
\end{equation}
 
For perfect reflection, 
i.e. for the momenta $k$'s such that ${\cal R}(k)=0$, 
one has $c=0$, $a=-b$ and from Eq.(\ref{eq_0}) $\psi_\alpha=-2a \sin{k}$.
Therefore  Eqs.(\ref{eq_alpha}) and (\ref{eq_altri}) reduce to
the eigenvalue equation for the adjacency matrix $A^0$ and,
apart from the trivial case $\cos(k)=0$, they are 
satisfied only if $2\cos{k}$ coincides with an eigenvalue of ${ A^0}$. 

At variance, in order to find the momenta $k$'s such that ${\cal T}(k)=0$ 
(perfect transmission), one has $b=0$, $a=c$ and 
from Eq.(\ref{eq_0}) $\psi_\alpha=0$. Therefore, Eqs.(\ref{eq_altri}) 
reduces to the eigenvalue equation for $A^r$ and it is satisfied only if 
$2\cos{k}$ coincides with an eigenvalue of $A^r$.

This general argument can be easily extended to the situation 
where $p$ identical
graphs $G^0$ are attached to $n=0$: indeed, now one has only to replace 
in Eq. (\ref{eq_0}) $\psi_{\alpha}$ with $p\psi_{\alpha}$ and the conditions
for ${\cal T}(k)=0$ and ${\cal R}(k)=0$ do not change.

The stated result holds for the case where the tunneling rates 
between the neighbour sites of $G^0$ are constant (and equal to 1 in the 
chosen units): 
however one can also consider in $G^0$ non-uniform tunneling rates  
$t^0_{\eta,\eta'}>0$, where 
$\eta$ and $\eta'$ are nearest-neighbour sites belonging to $G^0$ 
($t_{\eta,\eta'}^0=0$ if $A^0_{\eta,\eta'}=0$). The DNLSE  
at a site $\eta$ of $G^0$ becomes 
\begin{equation}
i \frac{\partial \psi_{\eta}}{\partial \tau}= - 
\frac{1}{2} \sum_{\eta'} t_{\eta,\eta'}^0 \psi_{\eta'}+\Lambda 
\mid \psi_\eta \mid^2 \psi_\eta
\end{equation} 
while the DNLSE at the sites $n$ of the of the chain remain unchanged. 
Now, the criterion states that ${\cal R}=1$ if 
$2\cos{k}$ coincides with an eigenvalue of the matrix $t_{\eta,\eta'}^0$, 
while ${\cal T}=1$ if $2\cos{k}$ is an eigenvalue of 
the matrix $t_{\eta,\eta'}^r$ defined as $t_{\eta,\eta'}^r=t_{\eta,\eta'}^0$ 
if $\eta$ and $\eta'$ belong to the reduced graph $G^r$.   

\section{Finite Linear Chains}

As a first simple application of the argument given in Section IV,  
we consider a single site $\alpha$ attached via a single link 
to the site $0$ of the unbranched chain. For Bose-Einstein 
condensates in optical lattices \cite{oberthaler03} 
the setup 
"infinite chain + single link" or "infinite chain + finite chain" 
may be realized by using two pairs of 
counterpropagating laser beams to create a star-shaped geometry 
in the $x$-$y$ plan \cite{brunelli04} and manipulating the frequencies of 
the superimposed harmonic magnetic potential 
so that in the $y$ direction only few sites 
can be occupied. In an optic context, 
chains of coupled waveguides are routinely built and studied 
\cite{kivshar03}: one
can obtain the configuration "infinite chain + single link" by
coupling a further waveguide to a waveguide of the chain. 

The DNLSE in the sites $0$ and $\alpha$ 
reads 
\begin{equation}
i \frac{\partial \psi_0}{\partial \tau}=-\frac{1}{2} (\psi_1+\psi_{-1}+
\psi_\alpha)+\Lambda \mid \psi_0 \mid^2 \psi_0 
\label{singolo_link_0}
\end{equation}
and 
\begin{equation}
i \frac{\partial \psi_\alpha}{\partial \tau}=-\frac{1}{2} \psi_0+
\Lambda \mid \psi_\alpha \mid^2 \psi_\alpha.
\label{singolo_link_alpha}
\end{equation}
It is transparent from Eqs.(\ref{singolo_link_0})-(\ref{singolo_link_alpha}) 
that the wavefunction $\psi_\alpha(\tau)$ may be interpreted as an 
additional local Fano degree of freedom, 
yielding the so-called Fano-Anderson 
model \cite{miroshnichenko03,flach03,miroshnichenko05}.  
In our approach, such degree of freedom is interpreted 
as a single link attached to the unbranched chain. As it is well known, 
the Fano-Anderson model describes interesting scattering properties: 
adding a generic finite graph (instead of a single link) 
gives rise to a yet richer variety of behaviors. We mention that 
in \cite{miroshnichenko05} the Fano degree of freedom is coupled 
to several sites of the unbranched chain: this would correspond in our 
description to a site linked to several sites of the chain, and, 
in general, to graphs attached to several sites of the chain. 
For simplicity, in the following we limit ourself to graphs inserted 
in a single site of the unbranched chain.

For large-fast solitons, when a single link is added to the unbranched 
chain, the 
reflection coefficient ${\cal R}$ from 
Eqs.(\ref{singolo_link_0})-(\ref{singolo_link_alpha}) 
is found to be \cite{miroshnichenko03}
\begin{equation}
{\cal R}=\frac{1}{1+4 \sin^2{(2k)}}:
\label{R_singolo_link}
\end{equation}
in the regime where $\tau_{disp} \gg \tau_{int}$, 
we verified that the numerical results of the soliton scattering against 
the link are in agreement with Eq.(\ref{R_singolo_link}) 
(see Fig.2). One sees that, when $k$ is approaching $\pi$ 
(solitons becoming slower), the agreement becomes worse.
From the general results, 
there are no fully transmitted momenta and ${\cal R}=1$ only 
for $k=\pi/2$ and $k=\pi$, as one can also see by a direct inspection of 
(\ref{R_singolo_link}). 

One can also attach a finite chain of length $L$ at the site $0$. 
In the linear approximation
for large-fast solitons (i.e., Eq.(\ref{DLS})), the solution
corresponding to a plane wave
coming from the left of the unbranched chain is $\psi_n=a e^{ikn}+be^{-ikn}$
for $n<0$ and $\psi_n=c e^{ikn}$ for $n>0$, while in the attached chain 
$\psi_{\alpha}=f e^{ik \alpha}+ge^{-ik \alpha}$ 
($\alpha=1,\cdots,L$ denotes the sites of the attached chain, 
and $\alpha=1$ is the site linked to $n=0$). 
Eq.(\ref{DLS}) for $n=0$, $\alpha=1$, and $\alpha=L-1$ yields 
respectively 
\begin{equation}
a+b=c=f+g
\label{links_1}
\end{equation}
\begin{equation}
-\frac{1}{2} (a e^{-i k} + b e^{-i k} + c e^{i k} + f e^{i k} + g^{-i k})=
\mu (a+b)
\label{links_2}
\end{equation}
and
\begin{equation}
-\frac{1}{2} (f e^{i k (L-1)} + g e^{-i k (L-1)})=
\mu (f e^{i k L} + g e^{-i k L})
\label{links_3}
\end{equation}
where $\mu=-\cos{k}$. 

We have five unknowns ($a$, $b$, $c$, $f$, and $g$) 
and four equations (\ref{links_1})-(\ref{links_2})
(the remaining condition being provided by the normalization). 
One may easily determine $b/a$, $c/a$, $f/a$, and $g/a$, getting  
\begin{equation}
\frac{b}{a}=\frac{e^{2 i k}(e^{2ikL}-1)}{1-2e^{2ik}+e^{2ik(L+2)}},
\end{equation}
which leads to
\begin{equation}
{\cal R}=\mid b/a \mid^2 = 
\frac{\sin^2{(kL)}}{[\cos{(kL)}-\cos{(k(L+2))}]^2+\sin^2{(kL)}}.
\label{R_L_links}
\end{equation}
and ${\cal T}=\mid c/a \mid^2 = 1 - {\cal R}$. 
For $L=1$, Eq.(\ref{R_L_links}) reduces to Eq.(\ref{R_singolo_link}). 
In agreement with the general argument of Section IV,
the number of minima and maxima increases with $L$.
Eq.(\ref{R_L_links}) for $L=2$ is compared in Fig.2 with the numerical 
results. 

We notice that in the limit $L \to \infty$ the considered problem 
corresponds to the propagation of a soliton in a the so-called 
{\em star graph}, which has been 
recently investigated in the context of two-dimensional networks of nonlinear 
waveguide arrays \cite{christodoulides01} and 
$Y$-junctions for matter waves \cite{kevrekidis03}.

\section{Cayley trees}
 
Eq.(\ref{R_L_links}) yields 
that the values of $k$ allowing for perfect 
reflection (${\cal R}=1$) for the length $L$ coincide with the  momenta of full
transmission (${\cal T}=1$) when $G^0$ is a chain of length $L+1$. This
property is  readily understood: in fact, if $G^0$ is
a chain of length $L$ the $k$'s for which ${\cal R}=1$ correspond to
the energy levels of $G^0$. If $G^0$ is a chain of length $L+1$, the values of
$k$'s for which ${\cal T}=1$ correspond to energy levels  of the reduced graph
$G^r$, which, in this case, is again given by a chain of length $L$.  This is
clearly  a general property of any self-similar graph. As an example, we study
in this section the situation  in which $G^0$ is a Cayley tree of branching
rate $p$ and generation $L$ (see Fig.3). 

Let us consider the linear 
approximation  for large-fast solitons  Eq.(\ref{DLS}). 
The plane wave coming from the left of the unbranched chain is 
$\psi_n=a e^{ikn}+be^{-ikn}$ 
for $n<0$ and $\psi_n=c e^{ikn}$ for $n>0$. This fixes
$\mu=-\cos{k}$. Furthermore, from the continuity in $0$ it follows $a+b=c$.

For a Cayley tree, the eigenfunction must have, 
by symmetry, the same value at all the sites 
belonging to the same generation. If we denote by $\psi_\beta$ 
the eigenfunction at the sites at distance 
$\beta=1,\cdots,L$ from $n=0$, 
the eigenvalue equation (\ref{DLS}) at the site $\alpha=2,\cdots,L-1$ reads
\begin{equation}
-\frac{1}{2}(\psi_{\alpha-1}+p\psi_{\alpha+1})=\mu \psi_{\alpha}.
\label{eig_eq_Cayley}
\end{equation}
The plane wave solutions of Eq.(\ref{eig_eq_Cayley}) can be written as 
\begin{equation}
\psi_{\alpha}=\frac{1}{p^{\alpha/2}}(f e^{ik' \alpha}+ge^{-ik' \alpha}),
\label{cayley}
\end{equation}
where $\alpha=1,\cdots,L$: in this way 
one gets from Eq.(\ref{eig_eq_Cayley}) 
$\mu=-\sqrt{p} \cos{k'}$, so that 
$k'=\arccos{(p^{-1/2} \cos{k})}$. 
Eq.(\ref{DLS}) for $n=0$, $\alpha=1$, and $\alpha=L-1$ gives respectively  
\begin{equation}
-\frac{1}{2} (a e^{-i k} + b e^{i k} + c e^{i k})-\frac{1}{2 \sqrt{p}}
(f e^{i k'} + g e^{-i k'})=\mu(a+b)
\label{cayley_1}
\end{equation}
\begin{equation}
-\frac{1}{2} (a + b)-\frac{1}{2}
(f e^{2 i k'} + g e^{-2 i k'})=\frac{\mu}{\sqrt{p}} 
(f e^{i k'} + g e^{-i k'})
\label{cayley_2}
\end{equation}
and 
\begin{equation}
-\frac{1}{2} (f e^{i k' (L-1)} + g e^{-i k' (L-1)})=
\frac{\mu}{\sqrt{p}} (f e^{i k' L} + g e^{-i k' L}).
\label{cayley_3}
\end{equation}
Using Eqs.(\ref{cayley_1})-(\ref{cayley_3}) and the condition $a+b=c$, 
one can determine $b/a$, $c/a$, $f/a$, and $g/a$: the resulting 
expressions is rather involved and here we will not explicitly write them. 
In Fig.3 we plot the numerical and analytical 
results for the reflection coefficient ${\cal R}$ when two Cayley trees, 
respectively with $L=5$ and $L=6$, are attached 
to the unbranched chain. One sees that when one pass to the next 
generation, the momenta for which full reflection occurs 
becomes momenta of full transmission.

\section{Further Examples of inserted graphs}

In general, it is possible to consider the scattering of a large-fast soliton 
through a large variety of inhomogeneous networks. Here,
we analyze three further examples of network topologies: 
stars, loops and complete graphs. In each case, 
the reflection coefficients for large-fast solitons are
derived with a procedure analogous to the one adopted
in the previous section for Cayley trees.
The analytical findings are compared with numerical results.

{\em Loops:} Let us consider a loop graph, i.e., a finite chain  
of $L$ sites $\alpha_1, \cdots, \alpha_L$ such that 
the sites $\alpha_1$ and $\alpha_L$ are linked to the site 
$n=0$ of the unbranched chain. For large-fast solitons,
the reflection coefficient ${\cal R}$ in the linear approximated regime is
\begin{equation}
{\cal R}=2 \, \, \, \frac{\left[ 1+\cos{(k(L-1))}\right]^2 +
\sin^2{(k(L-1))}}{6-2 \cos{(2k)}+5 \cos{k} \cos{(kL)}-\cos{(k(L+3))} 
+\sin{k} \sin{(kL)}}.
\label{R_L_loop}
\end{equation}
We note that the number of momenta of perfect reflection 
and perfect transmission increases with $L$. 
Furthermore, at large $L$, the transmission properties 
of the loops become similar to those of a finite (not closed) 
chain [see Eq.(\ref{R_L_links})]. 
In Fig.4 the numerical and analytical results are compared
and a figure of the loop graph with $L=3$ is provided.  

{\em Stars:} The $p$-star 
is the graph composed by a central site linked to $p$ 
sites, which are in turn connected only to the 
central site. Let us consider the case where $G^0$ is a $p$-star
graph and $\alpha$ is the center, linked to $n=0$. 
In the linear approximation we have
\begin{equation}
{\cal R}=\frac{1}{1+\left( \frac{\sin{(3k)}}{\cos{k}}-(p-1) \tan{k}\right)^2} 
\label{R_L_star}.
\end{equation}
In Fig.5 the numerical and analytical results are compared. 
Eq.(\ref{R_L_star}) shows that perfect transmission (${\cal R}=0$) 
is obtained only for for the momentum $k=\pi/2$. 
This can be directly proved
applying the criterion of Section IV: indeed 
$G^0$ is a $p$-star, while $G^r$ 
consists of $p$ disconnected sites. 
Therefore, all the eigenvalues of the adjacency matrix $A^r$ equal
zero, and the only momentum of perfect transmission is $k=\pi / 2$. 

{\em Complete graphs:} The complete graph $K_M$ of $M$ sites 
is the graph where every pair of sites in linked \cite{harary69}: e.g, 
$K_3$ is a triangle.
Inserting $K_{L+1}$ at the site $n=0$ of the unbranched chain 
(so that $n=0$ is one of the sites of $K_{L+1}$), one gets:
\begin{equation}
{\cal R}=\frac{1}{1+4 \frac{(L-1-2 \cos{k})^2}{L^2} \sin^2{k}}. 
\label{R_L_simpl}
\end{equation}
For $L >> 1$, ${\cal R} \approx 
1/ (1+4 \sin^2{k})$, therefore 
the complete graph behaves as a single effective 
defect [compare with Eq.(\ref{R_singolo_link})]. The comparison between the 
numerical and analytical results is presented in Fig.6.   

\section{Conclusions} 

As a first step in addressing the issue of the interplay 
between nonlinearity and topology, we studied the discrete nonlinear 
Schr\"odinger equation on a network 
built by attaching to a site of an unbranched chain a topological
perturbation $G^0$. 
The relevant situation corresponding to the Fano-Anderson model 
is obtained when one considers a single link attached to the linear chain. 
We showed that, by properly selecting the attached graph, 
one is able to control the perfect 
reflection and transmission of traveling solitons. 
We derived a general criterion 
yielding - once the energy levels of the graph $G^0$ is known - 
the momenta at which the soliton is fully reflected or fully transmitted. 
For self-similar graphs $G^0$, we found that the values 
of momenta for which perfect reflection occurs become perfect 
transmission momenta when the next generation of the graph is considered. 
For finite linear chains, loops, stars and complete graphs,  
we studied the transmission coefficients and we compared 
numerical results form the discrete nonlinear 
Schr\"odinger equation with analytical estimates.  
Our results evidence the remarkable influence of topology 
on nonlinear dynamics and are amenable to interesting applications 
in optics since one may think of engineering inhomogeneous chains acting 
as a filter for the motion of soliton \cite{burioni05}.

{\em Acknowledgments:} We thank M. J. Ablowitz, 
P. G. Kevrekidis and B. A. Malomed for discussions.

\begin{figure}[h]
\centerline{\psfig{%bbllx=5mm,bblly=6mm,bburx=184mm,bbury=137mm,%
figure=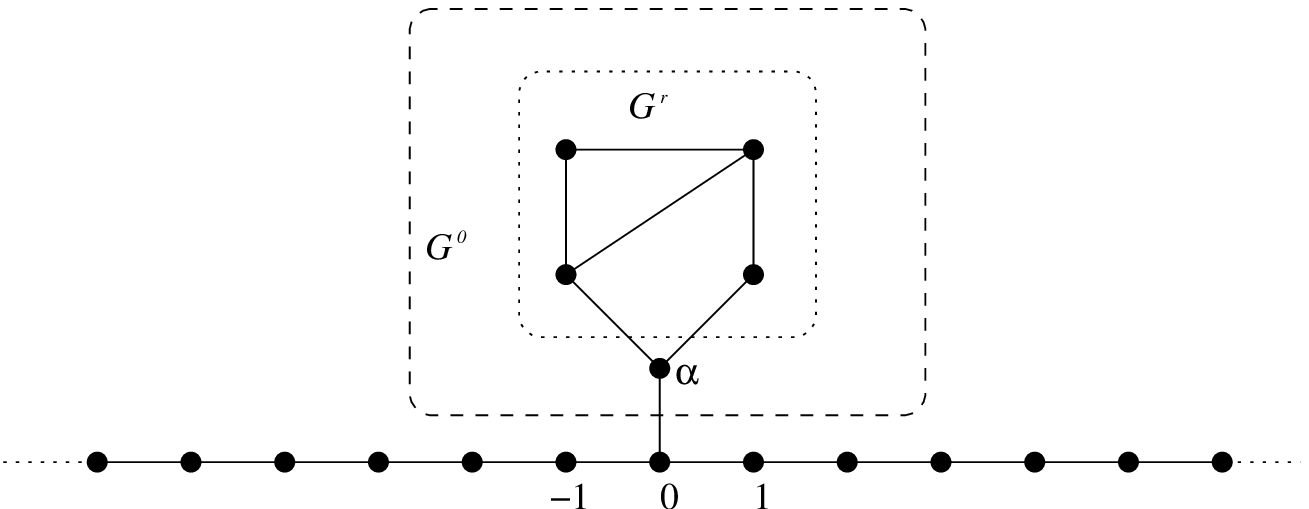,width=82mm,angle=0}}
\caption{Inserting a graph $G^0$ on a site of a linear chain: 
the points of the chain are denoted with integers $n$ and 
the point in which the graph is attached is $n=0$; 
$\alpha$ is the 
point of $G^0$ connected to $0$. 
$G^r$ is obtained subtracting $\alpha$ from $G^0$.}
\end{figure}

\begin{figure}[h]
\centerline{\psfig{%bbllx=5mm,bblly=6mm,bburx=184mm,bbury=137mm,%
figure=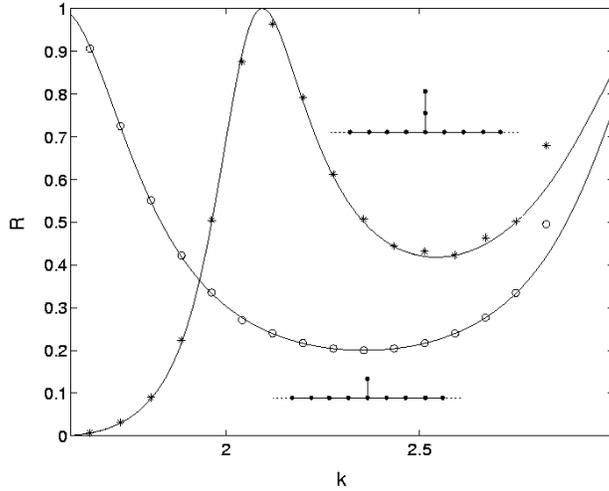,width=82mm,angle=0}}
\caption{Reflection coefficient $\cal R$ as a function of $k$ 
(with $k$ between $\pi/2$ and $\pi$) when a chain with length $1$ 
(i.e., a single link) and $2$ are attached. 
Empty circles ($L=1$) and stars ($L=2$) 
correspond to the numerical solution of Eq.(\ref{DNLS-gen}): 
in this figure, as well in the followings, 
as initial condition we choose a Gaussian with initial 
width $\gamma_0=40$ and momentum $k$. 
Solid lines correspond to the analytical prediction (\ref{R_L_links}).}
\end{figure}

\begin{figure}[h]
\centerline{\psfig{%bbllx=5mm,bblly=6mm,bburx=184mm,bbury=137mm,%
figure=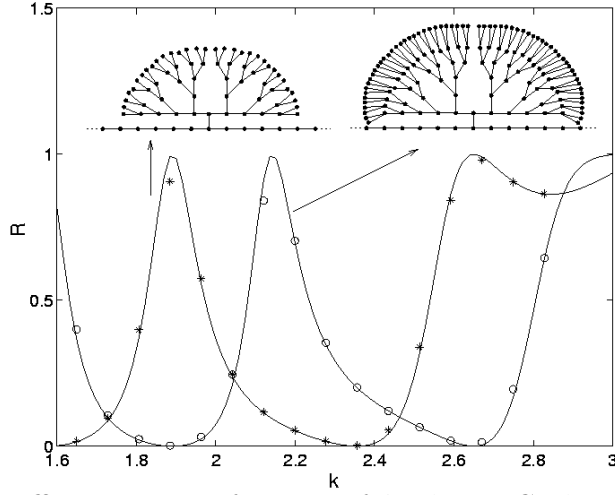,width=82mm,angle=0}}
\caption{Reflection coefficient $\cal R$ as a function of $k$ 
when a Cayley tree with length $5$ 
and $6$ are attached. 
Empty circles ($L=6$) and stars ($L=5$) 
correspond to the numerical solution of Eq.(\ref{DNLS-gen}). 
Solid lines correspond to the analytical prediction (see text). 
As required from the general argument in Section III, the values of $k$ 
for which one has perfect reflection (${\cal R}(k)=1$) for $L=5$ correspond 
to perfect transmission (${\cal R}(k)=0$) for $L=6$.}
\end{figure}

\begin{figure}[h]
\centerline{\psfig{%bbllx=5mm,bblly=6mm,bburx=184mm,bbury=137mm,%
figure=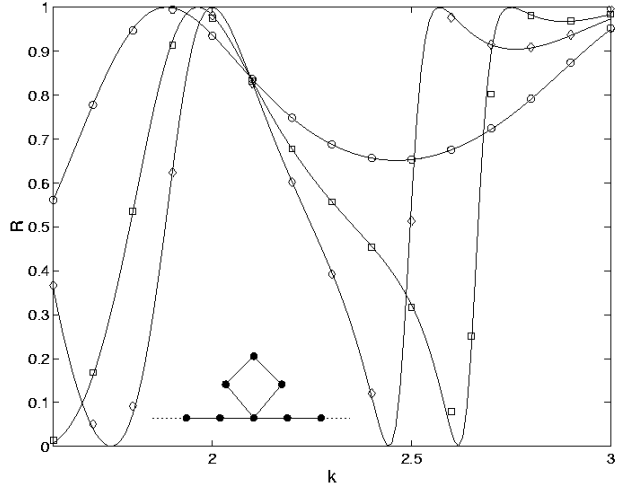,width=82mm}}
\caption{Reflection coefficient $\cal R$ as a function of $k$ 
when loops with $L=4$, $7$ and $10$ are inserted at a site 
of the unbranched chain.  
Empty circles ($L=4$), squares ($L=7$) and diamonds ($L=10$) are obtained from 
the numerical solution of Eq.(\ref{DNLS-gen}). 
Solid lines correspond to the analytical prediction (\ref{R_L_loop}).
The small figure represents a loop with 
$L=3$ inserted in the unbranched chain.}
\end{figure}

\begin{figure}[h]
\centerline{\psfig{%bbllx=5mm,bblly=6mm,bburx=184mm,bbury=137mm,%
figure=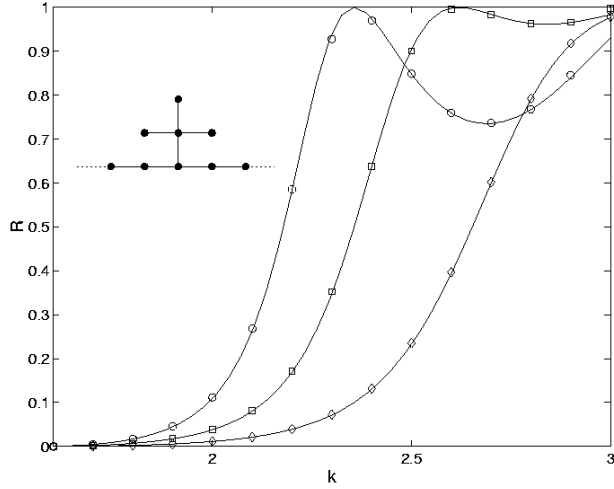,width=82mm}}
\caption{Reflection coefficient $\cal R$ as a function of $k$ 
when stars with $p=3$, $4$ and $6$ are inserted.
Empty circles ($p=2$), squares ($p=3$) and diamonds ($p=5$) are obtained from 
the numerical solution of Eq.(\ref{DNLS-gen}). 
Solid lines correspond to the analytical prediction (\ref{R_L_star}).
The inset represents the situation where the attached graph $G^0$ is
a star with $p=3$.}
\end{figure}

\begin{figure}[h]
\centerline{\psfig{%bbllx=5mm,bblly=6mm,bburx=184mm,bbury=137mm,%
figure=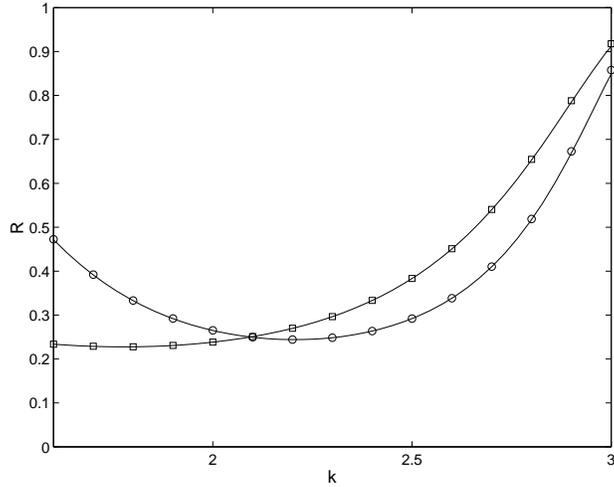,width=82mm}}
\caption{Reflection coefficient $\cal R$ as a function of $k$ 
when the complete graphs $K_3$ and $K_{11}$ are inserted at a site 
of an unbranched chain. Empty circles ($K_3$) and 
squares ($K_{11}$) are obtained from 
the numerical solution of Eq.(\ref{DNLS-gen}). 
Solid lines correspond to the analytical prediction (\ref{R_L_simpl}).  }
\end{figure}

%\end{multicols}

\end{document}